\documentclass{eptcs}
\providecommand{\event}{LINEARITY 2012}

\providecommand{\firstpage}{1}

\usepackage{url} 

\usepackage[utf8]{inputenc}
\usepackage[english]{babel}
\usepackage{cite}
\usepackage{amsmath,amssymb,stmaryrd}
\usepackage{mathrsfs}

\usepackage{algorithmic}
\usepackage{algorithm}

\usepackage{float}
\usepackage[all]{xy}
\entrymodifiers={+!!<0pt,\fontdimen22\textfont2>}

\newcommand{\Ra}{\Rightarrow}
\newcommand{\R}{\mathcal R}

\newcommand{\uop}[1]{\{ #1 \}}

\newcommand{\pp}[1]{\mathit{#1}}
\newcommand{\mathtab}{\;\;\;\;}
\newcommand{\oneto}[1]{\llbracket 1,#1\rrbracket}
\newcommand{\injection}{\overset{\textit{inj}}{\longrightarrow}}
\newcommand{\bijection}{\overset{\textit{bij}}{\longrightarrow}}
\newcommand{\restrict}[3]{#1_{#2\restriction #3}}

\usepackage{color}
\newcommand{\todo}[1]{}
\newcommand{\emphdef}[1]{\textbf{#1}}

\newcommand{\fov}[1]{\mathtt{#1}} 
\newcommand{\hov}[1]{\mathfrak{#1}} 
\newcommand{\ho}[1]{\widehat{#1}} 
\newcommand{\meta}[1]{\mathscr{#1}} 

\newcommand{\G}{\meta G}
\newcommand{\N}{\meta N}
\newcommand{\bP}{\meta P}

\newcommand{\Nats}{\mathbb N}
\newcommand{\Part}{\mathcal P}

\newcommand{\X}{\mathcal X}

\newcommand{\Land}{\wedge} 
\newcommand{\Lor}{\vee}

\newtheorem{defn}{Definition}

\newtheorem{thm}{Theorem}

\newcommand{\nAxn}{Ax}
\newcommand{\nWn}{W}
\newcommand{\nCn}{C}
\newcommand{\nToIn}{\Rightarrow_{\mathscr I}}
\newcommand{\nToIcn}{\Rightarrow_{\mathscr I}^{c}}
\newcommand{\nToEn}{\Rightarrow_{\mathscr E}}
\newcommand{\nAndIn}{\Land_{\mathscr I}}
\newcommand{\nAndEleftn}{\Land_{\mathscr E_l}}
\newcommand{\nAndErightn}{\Land_{\mathscr E_r}}
\newcommand{\nSn}{s}

\newcommand{\pAxp}{{p}}
\newcommand{\pAxin}{{in}}
\newcommand{\pWp}{{p}}
\newcommand{\pCp}{{p}}
\newcommand{\pCoutl}{{out_l}}
\newcommand{\pCoutr}{{out_r}}
\newcommand{\pToIp}{{p}}
\newcommand{\pToIinl}{{in_l}}
\newcommand{\pToIinr}{{in_r}}
\newcommand{\pToIs}{{s}}
\newcommand{\pToIcp}{{p}}
\newcommand{\pToIcinl}{{in_l}}
\newcommand{\pToIcinr}{{in_r}}
\newcommand{\pToEp}{{p}}
\newcommand{\pToEinl}{{in_l}}
\newcommand{\pToEinr}{{in_r}}
\newcommand{\pAndIp}{{p}}
\newcommand{\pAndIinl}{{in_l}}
\newcommand{\pAndIinr}{{in_r}}
\newcommand{\pAndEleftp}{{p}}
\newcommand{\pAndEleftin}{{in}}
\newcommand{\pAndErightp}{{p}}
\newcommand{\pAndErightin}{{in}}
\newcommand{\pSp}{{p}}
\newcommand{\pSin}{{in}}
\newcommand{\pSout}{{out}}

\newtheorem{example}{Example}

\title{Higher-order port-graph rewriting}
\author{
  Maribel Fern\'andez\institute{King's College London}\email{Maribel.Fernandez@kcl.ac.uk}
 \and
  S\'ebastien Maulat \institute{\'Ecole Normale Supérieure de Lyon}\email{Sebastien.Maulat@ens-lyon.fr}}
\date{}
\def\titlerunning{Higher-order port-graph rewriting}
\def\authorrunning{M. Fern\'andez and S. Maulat}

\begin{document}
\selectlanguage{english}

\maketitle{}

\begin{abstract}

The biologically inspired framework of port-graphs has been successfully used to specify complex systems. It is the basis of the PORGY modelling tool. To facilitate the specification of proof normalisation procedures via graph rewriting, in this paper we add higher-order features to the original port-graph syntax, along with a generalised notion of graph morphism.  We provide a matching algorithm which enables to implement higher-order port-graph rewriting in PORGY, thus one can visually study the dynamics of the systems modelled. We illustrate the expressive power of higher-order port-graphs with examples taken from proof-net reduction systems.

\end{abstract}

\section{Introduction}

 {\em Rewriting systems}~\cite{Plump98termgraph} are used to specify and study computational processes, where the execution of a program is described as a sequence of transformation steps on syntactic objects.
For instance, in {\em term rewriting}~\cite{BaaderNipkow_all_that},
objects are abstract syntax trees, and their rewriting consists of
replacing subtrees. In the  functional paradigm,
representing a program as a term enables, amongst other things, to
specify evaluation strategies, and to prove properties of
computations, such as termination.

Graphical formalisms are used in various fields of computer science,
and \emph{graph rewriting} provides visual support for studying the dynamics
of complex structures, such as proofs, programs or biological systems. Graph
rewriting rules describe graph transformations; a rewriting step
consists of replacing an instance of the left-hand side with the
right-hand side.

{\em Port-graphs}~\cite{Andrei08,AndreiK08c} are a specific class of
labelled graphs introduced as an abstract representation of proteins,
and used to model biochemical interactions and autonomous systems.
Port-graphs have also been used to study and visualise
the normalisation of proof nets~\cite{proof_graphs}. Port-graph rewriting 
has been implemented in the PORGY environment~\cite{PORGY_main}.

Although the original notion of port-graphs provides a natural
graphical representation of proofs encompassing proof nets and
interaction nets, as detailed in \cite{proof_graphs}, the associated
rewriting system suffers from two drawbacks. First, the
cut-elimination procedure cannot be expressed directly, and its
encoding involves a huge enumeration of cases. Second, the duplication
and erasure of subproofs during normalisation, performed
locally, leads to additional rules that are not directly linked with
proof theory.

\paragraph{Contribution.}
We address the problems mentioned above, by defining an
extension of the original port-graph rewriting notion with
higher-order features. The extension provides further functionalities
to program with port-graph rewriting rules, so that the encoding of
the proof normalisation procedure in intuitionistic logic can be
expressed in a simple and natural way. We illustrate this extension
through examples. A matching algorithm is provided, enabling the
automation of the associated rewriting relation.

In order to focus on this extension, we chose to base our work on a 
restriction of the preexisting (first-order) port-graphs. Nevertheless, 
we keep a sufficiently expressive part of the port-graph syntax to represent 
proofs as port-graphs as described in~\cite{proof_graphs}, and postulate that 
this extension can be generalised to include all the features of the original 
syntax defined in \cite{Andrei08,AndreiK08c}.

\paragraph{Related work.}
Higher-order extensions have been defined for more restricted
formalisms: Term rewriting has been extended with higher-order
features, with formalisms such as {\em Combinatory Reduction
  Systems}~\cite{KlopOR93} and {\em Nominal Rewriting
  Systems}~\cite{FernandezGM04} amongst others. Higher-order graph
rewriting theories have been defined in
\cite{fernandezMP07,LaneveC:phd} via textual calculi instead of
graphical formalisms.  The examples that motivate the higher-order
extension of port-graphs presented in this paper come from the
graphical representation of proofs in intuitionistic logic given
in~\cite{proof_graphs}. Other graphical formalisms for the
representation of proofs have been proposed
in~\cite{BussS:91,CarboneA:10,DeOliveira:phd} amongst others.

\paragraph{Organisation of the article.}
In Section~\ref{sec:prelim} we present some preliminary notions. In Section~\ref{sec:hopg} we define the syntax of higher-order port-graphs, and in Section~\ref{sec:matr} we give the associated rewriting calculus and apply it to some example port-graphs representing proofs. We provide a matching algorithm in Section~\ref{sec:impl} and discuss properties in Section~\ref{sec:prop}. Section~\ref{sec:conc} concludes.

\section{Preliminaries and motivation}
\label{sec:prelim}

\paragraph{Graphs from proofs.}

Structured formalisms such as graphs can be used to represent proofs in a simple and concise way. For instance, a graph based formalism generalising Lafont's interaction nets~\cite{LafontY:intn} is introduced in \cite{proof_graphs}, which enables to represent proofs and processes over proofs in a natural way. Namely, proof derivations from intuitionistic logic, expressed in natural deduction style, are inductively translated into \emph{port-graphs}, so that the normalisation of a proof is visually expressed as a step by step process consisting in applying transformation rules on port-graphs.

\paragraph{Port-graphs.}
The computational model of port-graphs was introduced to model biochemical processes, for example, protein interactions where two proteins connect via sites of chemical compatibility. Visually, a port-graph is a graph where edges are attached to nodes at points called \emph{ports}. Below we give a short and informal introduction to port graphs (see~\cite{Andrei08,AndreiK08c} for more details and examples).

Nodes and ports are labelled with {\em names}, and the association of ports to nodes is subject to a typing via a {\em p-signature}, that associates a finite set of port names to a node name. A $p$-signature can be extended with variable nodes, that might have variable port-names. An edge $((v,p),(v',p'))$ connects the node $v$ to the node $v'$ via the ports $p$ and $p'$. A port in a node might be associated to a state (for instance, active/inactive or principal/auxiliary) and similarly, nodes can have associated properties (like colour or shape that are used for visualisation purposes).
  A port graph can be considered as a labelled graph where ports are represented by nodes, and the nodes are only connected to their ports.
  As a consequence, expressivity results and properties of labelled graphs can be translated to port graphs.

\paragraph{Port-graph rewriting.}

Rules are pairs of port-graphs describing
transformations. Intuitively, a rule consists of two port-graphs $L$
and $R$ and a mapping between their ports, given by the \emph{rule
  interface}, which is graphically represented in an arrow node
separating $L$ and $R$.  A rule specifies how to transform an
occurrence of $L$ into $R$ inside a given port-graph.  The arrow node
has the following characteristics: for each port $p$ in $L$, to which
corresponds a non-empty set of ports $\{p_1,\ldots, p_n\}$ in $R$, the
arrow node has a unique port $r$ and the incident directed edges
$(p,r)$ and $(r,p_i)$, for all $i=1,\ldots,n$; all ports from $L$ that
are deleted in $R$ are connected to the {\em black hole} port of the
arrow node. 

The application of a rule for port-graphs is inspired by the standard
definition for graphs, using the ``double pushout approach''
(see~\cite{CorradiniMREHL97}), and relies on a definition of
matching. In this definition, applying a rule $L\to R$ on a graph $G$
is performed in four steps:
\begin{itemize}
\item find a matching $m$ from $L$ to $G$
\item define the context graph $G^{-} = G \backslash m(L)$
\item add $m(R)$ to $G^{-}$
\item reconnect $m(R)$ and $G^{-}$ as specified by the rule's interface
\end{itemize}


\section{Higher-order port-graphs}
\label{sec:hopg}
In this section we introduce higher-order port-graphs.
Throughout the paper we use the following conventions:
ordered pairs are written $(a,b)$;
unordered pairs are written $\uop{a,b}$, so $\uop{a,b}=\uop{b,a}$. 

\subsection{Typing}

The formal definition of higher-order port-graph relies on the association of port names to node names via a {\em p-signature}. 
\begin{defn}[$p$-signature]
  A \emphdef{$p$-signature} is a tuple of disjoint sets
  \[\nabla^\X_\G = ((\nabla_\N,\X_\N),\X_{\G},(\nabla_\bP,\X_\bP))\] together with two functions $\pp{arity}$ and $\pp{Interface}$ such that:
  \begin{itemize}
  \item $(\nabla_\N,\X_\N)$ are the sets of constant and variable first-order node names
  \item $\X_\G$ is the set of variable higher-order node names 
  \item $(\nabla_\bP,\X_\bP)$ are the sets of constant and variable port names
  \item $\pp{arity}: \nabla_\N \cup \X_\N \cup \X_\G \to \Nats$ associates a number of ports to a node name
  \item $\pp{Interface}$ associates the names of its ports to any node name:
    \begin{itemize}
    \item $\forall N\in\nabla_\N, \; \pp{Interface}_N: \oneto{\pp{arity}(N)} \injection \nabla_\bP$
    \item $\forall X\in\X_\N, \; \pp{Interface}_X: \oneto{\pp{arity}(X)} \injection \nabla_\bP \cup \X_\bP$
    \item $\forall \hov{X}\in\X_\G, \; \pp{Interface}_{\hov{X}}: \oneto{\pp{arity}(\hov{X})} \injection \X_\bP$
    \end{itemize}
  \end{itemize}
\end{defn}

Note that the interface of a constant first-order node name is a list of pairwise different constant port names. The interface of a variable first-order node name may contain variables, and the interface of a higher-order node name is a list of variables.

In our examples, the following symbols will be used:
\begin{align*}
  \fov{a},\fov{b},\fov{c},\hdots \in\nabla_\bP && \fov{A},\fov{B},\fov{C},\hdots \in\nabla_\N \\
  \fov{x},\fov{y},\fov{z},\hdots \in\X_\bP && \fov{U},\fov{V},\fov{W},\hdots \in\X_\N && \hov{X},\hov{Y},\hov{Z},\hdots \in\X_\G
\end{align*}

As shown in Figure~\ref{fig:sig_dummy}, a signature can be represented by either a table associating to any node name its arity and the list of its ports' names, or by a graph representing disconnected nodes with names and port names. Even if a node is disconnected, the location of its ports are indicated by ``dangling edges''. As the port names are unique, one can forget to represent the physical identifiers of ports on a node.

\begin{figure}[H]
  \begin{tabular}{cc}
 \begin{small}
  $\begin{array}{c||c||c|c|c}
    \begin{array}{c} \text{node}\\\text{name} \end{array} &
    \text{arity} & 
    \begin{array}{c} 1^{st} \text{ port}\\ \text{name} \end{array} &
    \begin{array}{c} 2^{nd} \text{ port} \\ \text{name} \end{array} &
    \dots \\ \hline
    \fov A & 1 & a & \\
    \fov B & 1 & b & \\
    \fov X & 1 & x & \\
    \hov X & 2 & y & z
   \end{array}$
 \end{small}
   &
\begin{small}
  $\begin{array}{c c}
    
    \begin{xy}
      \POS (0,0)*[o]=(5,5){\fov A}="A"*\frm{o}
      \POS "A"!U(1.35)="Aa"
      \POS "Aa"+(-1.5,.5) *{a}
      \POS "Aa"; "Aa"+(0,2) **\crv{}
    \end{xy}
    &
    \begin{xy}
      \POS (0,0)*[o]=(5,5){\fov B}="B"*\frm{o}
      \POS "B"!U(1.35)="Bb"
      \POS "Bb"+(-1.5,.5) *{b}
      \POS "Bb"; "Bb"+(0,2) **\crv{}
    \end{xy}
    \\ \\
    \begin{xy}
      \POS (0,0)*[o]=(5,5){\fov X}="X"*\frm{o}
      \POS "X"!U(1.35)="Xx"
      \POS "Xx"+(-1.5,.5) *{x}
      \POS "Xx"; "Xx"+(0,2) **\crv{}
    \end{xy}
    &
    \begin{xy}
      \POS (0,0)*[]=(10,5){\hov X}*+\frm{-}="hoX"
      \POS "hoX"!U(1.35)!L(.5)="hoXy"
      \POS "hoX"!U(1.35)!R(.5)="hoXz"
      \POS "hoXy"+(-2,1)*{y}
      \POS "hoXz"+(2,1)*{z}
      \POS "hoXy"; "hoXy"+(0,5) **\crv{}
      \POS "hoXz"; "hoXz"+(0,5) **\crv{}
    \end{xy}
    
  \end{array}$
\end{small}
  \end{tabular}
  \caption{A simple $p$-signature represented in two different ways.}
  \label{fig:sig_dummy}
\end{figure}

We illustrate the idea with an example from logic.

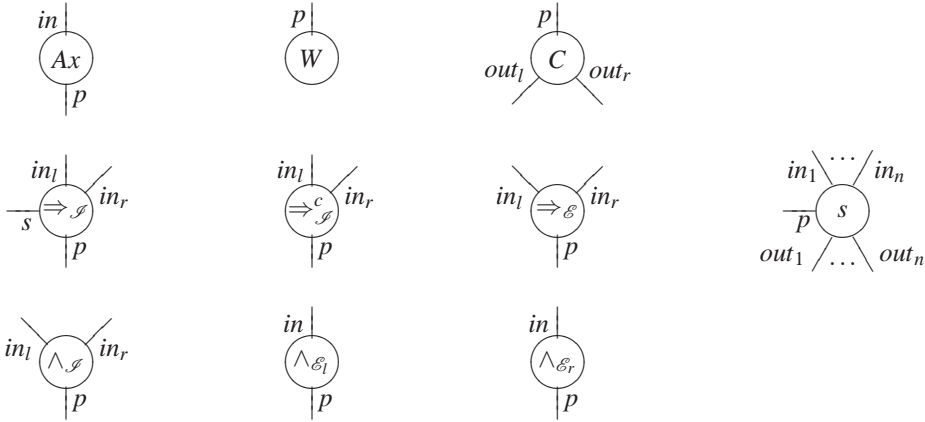
\begin{figure}[H]
\begin{small}

  \begin{tabular}[c]{ c @{\hspace{3pc}}  c @{\hspace{3pc}} c @{\hspace{4pc}} c}
    \begin{xy}
      \POS (0,0)*[o]=<20pt>{\nAxn}="Ax"*\frm{o}    
      \POS "Ax"!D(1.5)!R(0.5)*{\pAxp}              
      \POS "Ax"!U(1.5)!L(0.7)*{\pAxin}             
      \POS "Ax"!D(1.3) ; (0,-7) **\crv{}           
      \POS "Ax"!U(1.3) ; (0,7) **\crv{}            
      \POS (-10,-8)*{}; (10,8)*{}
    \end{xy}
    &
    \begin{xy}
      \POS (0,0)*[o]=<20pt>{\nWn}="W"*\frm{o}                    
      \POS "W"!U(1.5)!L(0.5)*{\pWp}                  
      \POS "W"!U(1.3) ;(0,7) **\crv{}
      \POS (-10,-8)*{}; (10,8)*{}
    \end{xy}
    &
    \begin{xy}
      \POS (0,0)*[o]=<20pt>{\nCn}="C"*\frm{o}                    
      \POS "C"!U(1.5)!L(0.5)*{\pCp}                  
      \POS "C"!L(2)!D(.5) *{\pCoutl}               
      \POS "C"!R(2)!D(.5) *{\pCoutr}               
      \POS "C"!U(1.3) ; (0,7) **\crv{}
      \POS "C" ; (-6,-6) **\crv{}
      \POS "C" ; (6,-6) **\crv{}
      \POS (-10,-8)*{}; (10,8)*{}
    \end{xy}
    
    \\
    \\

    \begin{xy}
      \POS (0,0)*[o]=<20pt>{\nToIn}="ToI"*\frm{o}                
      \POS "ToI"!D(1.5)!R(.5)*{\pToIp}              
      \POS "ToI"!U(1.5)!L(.8) *{\pToIinl}           
      \POS "ToI"!R(1.8)!U(.5) *{\pToIinr}            
      \POS "ToI"!L(1.5)!D(.5) *{\pToIs}              
      \POS "ToI"!D(1.3) ; (0,-7) **\crv{}
      \POS "ToI"!U(1.3) ; (0,7) **\crv{}
      \POS "ToI" ; (6,6) **\crv{}
      \POS "ToI"!L(1.3) ; (-7,0) **\crv{}
      \POS (-10,-8)*{}; (10,8)*{}
    \end{xy}
    &
    \begin{xy}
      \POS (0,0)*[o]=<20pt>{\nToIcn}="ToIc"*\frm{o}         
      \POS "ToIc"!D(1.5)!R(0.5)*{\pToIcp}       
      \POS "ToIc"!U(1.5)!L(0.8) *{\pToIcinl}    
      \POS "ToIc"!R(1.8)!U(.5) *{\pToIcinr}     
      \POS "ToIc"!D(1.3) ; (0,-7) **\crv{}
      \POS "ToIc"!U(1.3) ; (0,7) **\crv{}
      \POS "ToIc" ; (6,6) **\crv{}
      \POS (-10,-8)*{}; (10,8)*{}
    \end{xy}
    &
    \begin{xy}
      \POS (0,0)*[o]=<20pt>{\nToEn}="ToE"*\frm{o}                
      \POS "ToE"!D(1.5)!R(0.5)*{\pToEp}              
      \POS "ToE"!U(.5)!L(1.8) *{\pToEinl}            
      \POS "ToE"!U(.5)!R(1.8) *{\pToEinr}            
      \POS "ToE"!D(1.3) ; (0,-7) **\crv{}
      \POS "ToE" ; (-6,6) **\crv{}
      \POS "ToE" ; (6,6) **\crv{}
      \POS (-10,-8)*{}; (10,8)*{}
    \end{xy}
    &
    \begin{xy}
      \POS (0,0)*[o]=<20pt>{\nSn}="S"*\frm{o}

      \POS "S"!UL(1.5) *{\pSin_1}
      \POS "S"!U(1)!L(.4) ; (-4,8) **\crv{}
      \POS "S"!U(2) *{\dots}
      \POS "S"!UR(1.5)+(1,0) *{\pSin_n}
      \POS "S"!U(1)!R(.4) ; (4,8) **\crv{}
      \POS "S"!L(1.5)!D(.5) *{\pSp}
      \POS "S"!L(1.3) ; (-7,0) **\crv{}
      \POS "S"!DL(1.7)-(2,0) *{\pSout_1}
      \POS "S"!D(1)!R(.4) ; (4,-8) **\crv{}
      \POS "S"!D(2) *{\dots}
      \POS "S"!DR(1.7)+(2,0) *{\pSout_n}
      \POS "S"!D(1)!L(.4) ; (-4,-8) **\crv{}

      \POS (-10,-8)*{}; (10,8)*{}
    \end{xy}

    \\
    \\

    \begin{xy}
      \POS (0,0)*[o]=<20pt>{\nAndIn}="AndI"*\frm{o}              
      \POS "AndI"!D(1.5)!R(0.5)*{\pAndIp}            
      \POS "AndI"!U(.5)!L(1.8) *{\pAndIinl}          
      \POS "AndI"!U(.5)!R(1.8) *{\pAndIinr}          
      \POS "AndI"!D(1.3) ; (0,-7) **\crv{}
      \POS "AndI" ; (-6,6) **\crv{}
      \POS "AndI" ; (6,6) **\crv{}
      \POS (-10,-8)*{}; (10,8)*{}
    \end{xy}
    &
    \begin{xy}
      \POS (0,0)*[o]=<20pt>{\nAndEleftn}="AndEleft"*\frm{o}      
      \POS "AndEleft"!D(1.5)!R(0.5)*{\pAndEleftp}    
      \POS "AndEleft"!U(1.5)!L(0.7)*{\pAndEleftin}   
      \POS "AndEleft"!D(1.3) ; (0,-7) **\crv{}
      \POS "AndEleft"!U(1.3) ; (0,7) **\crv{}
      \POS (-10,-8)*{}; (10,8)*{}
    \end{xy}
    &
    \begin{xy}
      \POS (0,0)*[o]=<20pt>{\nAndErightn}="AndEright"*\frm{o}    
      \POS "AndEright"!D(1.5)!R(0.5)*{\pAndErightp}  
      \POS "AndEright"!U(1.5)!L(0.7)*{\pAndErightin} 
      \POS "AndEright"!D(1.3) ; (0,-7) **\crv{}
      \POS "AndEright"!U(1.3) ; (0,7) **\crv{}
      \POS (-10,-8)*{}; (10,8)*{}
    \end{xy}
  \end{tabular}
\end{small}
  \caption{A $p$-signature to represent proofs as port-graphs.}
  \label{fig:sig_proofs(graph)}
\end{figure}

\begin{example}
  The port-graph representation of proofs given in~\cite{proof_graphs} relies on the signature presented in Figure~\ref{fig:sig_proofs(graph)}: there are constant node names representing axioms, weakening and contraction rules, and the introduction and elimination rules for the connectives of the logic.
The node name $s$ is used to represent the scope of $\Rightarrow_{\!\cal I}$ (box).
\end{example}

\subsection{Syntax}

The definition of port-graphs (Definition~\ref{def:port-graph}) has
been designed to highlight the differences between first-order and
higher-order nodes. The ``hat'' notation (as in
``$\widehat{\text{hat}}$'') is used to distinguish the higher-order
functions and entities from the first-order ones.

The sets $V$ and $\ho{V}$ provide unique identifiers for first and higher-order nodes, and each node is labelled by a \emph{name} from a given $p$-signature. First-order nodes have first-order names and higher-order nodes have higher-order names. The name of a node determines the number of ports it has, which enables us to identify them concretely by integers starting from $1$. The signature then fixes the associated list of port names, that has no repetition.

An edge connects two nodes via their ports. Each port accepts at most one edge, but this assumption could be relaxed later -- for example by considering a maximal number of connections for each port, depending on its name.

\begin{defn}[Port graph]
\label{def:port-graph}
  A \emphdef{labelled higher-order port-graph} over the p-signature $\nabla^\X_\G$ is a tuple composed of:
  \begin{itemize}
  \item $V$ and $\ho{V}$ are finite sets of first-order and higher-order nodes, respectively.
  \item $lv: V \to \nabla_\N \cup \X_\N$ and $\ho{lv}: \ho{V} \to \X_\G$ 
    are labelling functions associating first-order names to first-order nodes, and higher-order names to higher-order nodes. These two functions fully determine concrete properties of the nodes:
    \begin{itemize}
    \item $\pp{degree}: V\cup \ho V \to \Nats$ associates to every first-order or higher-order node its number of ports, which must coincide with the arity of its label:
      \begin{align*}
        \forall v\in V, \pp{degree}(v)=\pp{arity}(lv(v)) \\
        \forall \hov v\in \ho V, \pp{degree}(\hov v)=\pp{arity}(\ho{lv}(\hov v))
      \end{align*}
    \item $\forall v\in V, \; lp_v: \oneto{\pp{degree}(v)} \to \nabla_\bP\cup \X_\bP$ and $\forall \hov{v}\in \ho{V}, \;\; \ho{lp}_{\hov{v}}: \oneto{\pp{degree}(\hov v)} \to \X_\bP$ associate a port name to a port identifier: 
      \begin{align*}
        \forall v\in V, lp_v = \pp{Interface}_{lv(v)} \\
        \forall \hov v\in \ho V, \ho{lp}_{\hov{v}} = \pp{Interface}_{\ho{lv}(\hov{v})}
      \end{align*}
    \end{itemize}
  \item $E$ is a finite set of undirected edges between ports:
    \[E \subseteq \left\{
      \begin{array}{lrl}
        \multicolumn{2}{c}{((v_1,p_1),(v_2,p_2))} & \mid \\
        & (v_i,p_i) \in & (V\times \oneto{\pp{degree}(v_i)}) \\
        && \cup (\ho{V}\times \oneto{\pp{degree}(v_i)})
      \end{array}
    \right\}\]
  and to simplify, we assume that each specific port $(v,p)$ occurs at most once in $E$ (this is always the case in interaction nets).
  \end{itemize}
\end{defn}

When using several port-graphs, indices will be used to identify the corresponding sets and functions. For instance, the tuple $(V_G,\ho V_G,\dots)$ is associated to the port-graph $G$. \todo{A physical port on a node is written $v.i$ where $v\in V\cup \ho V$ is the unique identifier of the node and $i$ either the unique identifier of the port inside $v$ or its name (which happens to be also unique with our current definition)} 
The set of port-graphs over the $p$-signature $\nabla^\X_\G$ is denoted by $\G(\nabla^\X_\G)$ (or simply $\G$ when there is no ambiguity).

\begin{figure}[H]
  \input{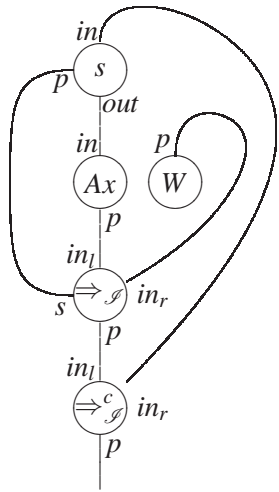}
  \caption{A port-graph representing a proof~\cite{proof_graphs}.}
  \label{fig:ex1_original}
\end{figure}

Intuitionistic proofs in natural deduction can be encoded as first-order port-graphs~\cite{proof_graphs}.
The proof of a sequent $\Gamma\vdash P$ is encoded as a port-graph with $\#(\Gamma) +1$ free ports corresponding to the premisses  and conclusion formulas. Each application of a rule is represented using nodes from the signature in Figure~\ref{fig:sig_proofs(graph)}. For instance, the port-graph representing the axiom inference rule ($A\vdash A$) is a simple node with two ports; Figure~\ref{fig:ex1_original} shows a proof of $\vdash A \Rightarrow B \Rightarrow A$ (nodes and port identifiers are usually omitted when representing graphically a port-graph). Details of this encoding are presented in \cite{proof_graphs}.

\section{Matching and rewriting}
\label{sec:matr}

\subsection{Sub-graph and Equality}
Two notions, which are trivial instances of morphisms, are introduced below. Intuitively, a {\em port sub-graph} is a subset of nodes, along with a subset of edges connecting them. Two port-graphs are considered {\em equal} when they are identical up to renaming of the concrete identifiers.

\begin{defn}[Port sub-graph]
  Given two port-graphs $G$ and $H$ over the same $p$-signature $\nabla^\X_\G$, $G$ is a \emphdef{port sub-graph} of $H$ if:\\
  \[
  \begin{array}{cc}
    V_G \subseteq V_H & \mathtab \ho V_G \subseteq \ho V_H \\
    lv_G = \restrict{lv}{H}{V_G} & \mathtab \ho{lv}_G = \restrict{\ho{lv}}{H}{\ho V_G} \\
    \multicolumn{2}{c}{E_G \subseteq E_H}
  \end{array}
  \]
\end{defn}

Note that since $H$ is typed, this definition implies $\pp{degree}_G = \restrict{\pp{degree}}{H}{V_G\cup \ho V_G}$, as well as $\forall v, lp_{G,v} = lp_{H,v}$ and $\forall \hov v, \ho{lp}_{G,\hov v} = \ho{lp}_{H,\hov v}$.

\begin{defn}[Equality]
  Two port-graphs $G$ and $H$ over the same signature are \emphdef{syntactically equal} via $(tr,\ho{tr})$ when $tr$ and $\ho{tr}$ are two bijections:\\
  \[(tr:V_G \bijection V_H,\ho{tr}:\ho{V}_G \bijection \ho{V}_H)\]
  such that:
  \begin{itemize}
  \item $lv_H=lv_G\circ tr^{-1}$ and $\ho{lv}_H= \ho{lv}_G \circ \ho{tr}^{-1}$
  \item 
    $E_H = \left\{
      \uop{(v_1,p_1),(v_2,p_2)} \mid
      \begin{matrix}
        \uop{(tr^{-1}(v_1),p_1),(tr^{-1}(v_2),p_2)} \in E_G \\
        \Lor \uop{(\ho{tr}^{-1}(v_1),p_1),(tr^{-1}(v_2),p_2)} \in E_G \\
        \Lor \uop{(\ho{tr}^{-1}(v_1),p_1),(\ho{tr}^{-1}(v_2),p_2)} \in E_G
      \end{matrix}
    \right\}
    $
  \end{itemize}
\end{defn}

Again, in $H$, the preservation of node names implies a preservation of the list of ports for each node. 

A \emph{full port sub-graph} is a sub-graph containing all edges between the selected nodes. A full port sub-graph $G$ of $H$ can be seen as a subset of nodes of $H$, with the same names and port lists, and with all the edges that link them to each other. Checking this property is purely syntactic, so easy to implement.

\subsection{Matching}

A definition of matching can now be given, using a notion of morphism. Intuitively, a morphism relates the elements of two port-graphs $G$ and $H$ in instantiating $G$ in a sub-graph of $H$. This is performed by mapping first-order nodes to first-order nodes, higher-order nodes to port-graphs, and edges to edges, while preserving the interface and connections between ports.

A definition of \emph{interface} has to be given for a port-graph, in order to formalise these preservation constraints.
The interface of a port-graph is the set of all the free ports it has. As in the case of simple nodes, the interface represents the points through which it can connect to the outside.

\begin{defn}[Interface of a port-graph]
  The interface $\pp{Interface}_\G(G)$ of a port-graph $G$ is the set of its ports $(v,p)$ that are not connected (i.e., that do not appear in $E_G$).
\end{defn}

 The definition of morphism should be as restrictive as possible --- without hampering the expressivity regarding proofs --- in order to decrease the number of morphisms between two port-graphs. The aim is to help preventing a combinatorial explosion, thus enabling a simple and efficient implementation.

Regarding its first-order part, a morphism instantiates or renames each variable node with a function $\sigma_\N$. The name of a first-order node is preserved if it is constant, and translated by $\sigma_\N$ if it is variable. Due to typing, an image node has the same number of ports as its antecedent, and these ports are in a one-to-one correspondence that preserves constant port names.
A higher-order node $\hov v$ is mapped to a sub-graph of $H$. The interface of $\hov v$ is bijectively mapped to the interface of its image.
Finally, these two mappings provide an injective translation of ports, such that sources and targets of edges are preserved.

\begin{defn}[Morphism]
  Given two port-graphs $G$ and $H$ over the same p-signature $\nabla^\X_\G$, a (higher-order) \emphdef{port-graph morphism} is a triple of functions
  \[ f=( f_V:V_G\to V_H,
  f_{\ho{V}}: \ho{V}_G\to \G(\nabla^\X_\G),
  f_E:E_G\to E_H) \]
 relating the \emph{pattern} $G$ and $H$, and satisfying the following properties:
  \begin{itemize}
  \item \emphdef{Instantiation of first-order variables}\\
    there exists a mapping for first-order variable nodes:
    \[ \sigma_\N: \X_\N \to \nabla_\N \cup \X_\N \]
    such that:
    \begin{itemize}
    \item constant node names are preserved, and $\sigma_\N$ instantiates or renames first-order variable nodes:\\
      $\forall v\in V_G,
      lv_H(f_V(v)) =
      \begin{cases}
        lv_G(v) & \mbox{if } \; lv_G(v)\in\nabla_\N\\
        \sigma_\N(lv_G(v)) & \mbox{if } \; lv_G(v)\in\X_\N
      \end{cases}$
    \item $\sigma_\N$ specifies and renames ports:\\
      \[\begin{array}{lcl}
        \forall X\in \X_\N, & & \pp{arity}(\sigma_\N(X)) = \pp{arity}(X) \\
        & \Land & \forall 1\leqslant p \leqslant \pp{arity}(X),\quad \pp{Interface}_X(p)=n\in \nabla_\bP \Rightarrow \pp{Interface}_{(\sigma_\N(X))}(p)=n
      \end{array}\]
    \end{itemize}
  \item \emphdef{Instantiation of higher-order variables}\\
    for each higher-order variable $\hov{X} \in {\ho{\X}}$, there exists
    \begin{itemize}
    \item a port-graph $J_{\hov X}=(V_{\hov X},\ho{V}_{\hov X},lv_{\hov X},\hdots,E_{\hov X})$ over $\nabla^\X_\G$
    \item a bijection $\pp{tr\_ports}_{\hov X}:\oneto{\pp{arity}(\hov X)} \bijection \pp{Interface}(J_{\hov X})$
    \end{itemize}
    such that for all $\hov{v} \in \ho{V}_G$, let $\hov X =\ho{lv}_G(\hov v) $:
    \begin{itemize}
    \item $f_{\ho{V}}(\hov{v})$ is a full port sub-graph of $H$, and syntactically equal to $J_{\hov X}$ for $(tr,\ho{tr})$
    \end{itemize}
    we denote by $\pp{tr\_ports}_{\hov v}$ the bijective mapping of higher-order interface induced by $\pp{tr\_ports}_{\hov X}$ and $(tr,\ho{tr})$ as follows:
    \begin{itemize}
    \item $\pp{tr\_ports}_{\hov v}: \oneto{\pp{degree}_G(\hov v)} \bijection \pp{Interface}_\G(f_{\ho{V}}(\hov{v}))$
    \item $ \forall p\in\oneto{\pp{degree}_G(\hov v)},\\
      \pp{tr\_ports}_{\hov v}(p) =
      \begin{cases}
        (tr^{-1}(v'),i) & \mbox{if } \pp{tr\_ports}_{\hov X}(p)=(v',i) \mbox{ with } v'\in V_G \\
        (\ho{tr}^{-1}(\hov v'),i) & \mbox{if } \pp{tr\_ports}_{\hov X}(p)=(\hov v',i) \mbox{ with } \hov v'\in\ho V_G
      \end{cases}
      $
    \end{itemize}
  \item \emphdef{Injection}\\
    all the nodes in the images are disjoint:\\
    $\forall v\neq v', \forall \hov v\neq \hov v',
    \left\{
      \begin{matrix}
        f_V(v) \neq f_V(v')\\
        f_V(v) \notin f_{\ho V}(\hov v)\\
        (V_{( f_{\ho V}(\hov v) )} \cup \ho V_{( f_{\ho V}(\hov v) )}) \cap
        (V_{( f_{\ho V}(\hov v') )} \cup \ho V_{( f_{\ho V}(\hov v') )}) = \emptyset
      \end{matrix}
    \right.
    $
  \item \emphdef{Edge preservation}\\
    sources and targets of edges are preserved:
    \[
    \begin{array}{c}
      \forall e=\uop{(v_1,p_1),(v_2,p_2)}\in E_G , f_E(e)=\uop{(v_1',p_1'),(v_2',p_2')}\\
      \mbox{where } \forall i, (v_i',p_i') =
      \begin{cases}
        (f_V(v_i),p_i) & \mbox{if } v_i\in V_G\\
        \pp{tr\_ports}_{v_i}(p_i) & \mbox{if } v_i\in \ho V_G
      \end{cases}
    \end{array}
    \]
  \end{itemize}
If there is a higher-order port-graph morphism between $G$ and $H$ we say that they \emph{match}.
\end{defn}

\begin{figure}[h]
  \input{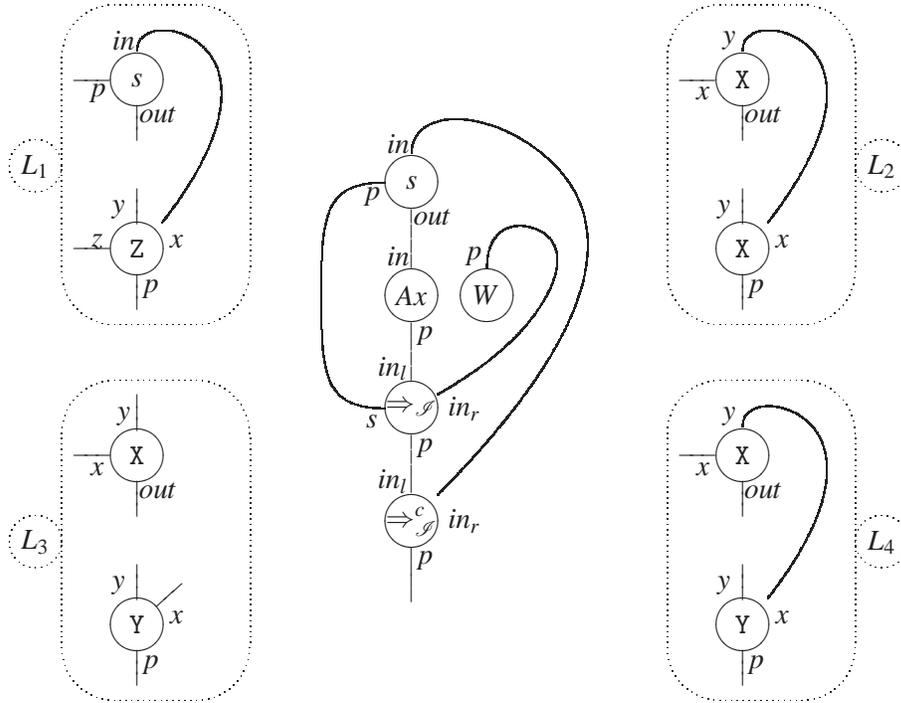}
  \caption{A target port-graph $G$ with four patterns port-graphs $L_1,\hdots,L_4$.}
  \label{fig:morphism_fo}
\end{figure}

\todo{The first-order part of this definition is illustrated in Figure~\ref{fig:morphism_fo}, where examples of pattern proof port-graphs are given - representing possible left hand-sides of rules - \dots. from In finding out which morphisms exist between the patterns and the central graph presented in Figure~\ref{fig:morphism_fo}, 
}
We now give some examples to illustrate this definition. Figure~\ref{fig:morphism_fo} shows four pattern port-graphs $L_1,\ldots,L_4$ and a central port-graph $G$. 

\begin{itemize}
\item There is no morphism from $L_1$ to $G$: By preservation of constant node names, the node $\nSn$ of $L_1$ would be mapped to the node $\nSn$ of $G$. Then, by preservation of the edges sources and targets, the image of the edge in $L_1$ would have an endpoint at the port $\pSin$ of $\nSn$ in $G$. By preservation of source and targets again, the node $\fov Z$ would be mapped to the node $\nToIcn$. But as they have different names and numbers of ports, this contradicts the definition.
\item There is no morphism from $L_2$ to $G$ either. Otherwise, by instantiation of first-order variables, the two physical nodes would be mapped to nodes with the same name. As all nodes have different names in $G$, the two image nodes would be physically identical, which contradicts the injection property of the morphism.
\item $L_3$ matches $G$. By conservation of the number of ports, a morphism from $L_3$ to $G$ maps the two nodes to $\nSn$ and $\nToIcn$. 
\item Similarly, a morphism between $L_4$ and $G$ maps $\fov X$ and $\fov Y$ to $\nSn$ and $\nToIcn$.
Note that the port variables are local to a node (more precisely, they are local to a node name but global to all the physical nodes that share this name).
\end{itemize}

\subsection{Rewriting}

A set of port-graph rewriting rules induces a rewriting relation, using the definition of morphism. We show in Section~\ref{sec:prop} that higher-order variables are more permissive than first-order ones (they allow us to express families of rules in a more concise way).

The notion of morphism induces a definition of matching: the {\em pattern} port-graph $L$ {\em matches} the {\em subject} port-graph $G$ if there exists a morphism $m$ from $L$ to $G$. This is denoted by $L \ll G$.

 The same operations are performed to define a {\em rewrite step} as in the case of graph rewriting (see Section~\ref{sec:prelim}). The rule interface (represented graphically in the arrow node) specifies the correspondence between ports in the interface of the left-hand side and ports in the interface of the right-hand side. Once the instantiation (via a morphism $m$) of the left hand side $L$ of a rule has been replaced in $G$ by the corresponding right-hand side $R$, the original edges between $G\backslash m(L)$ and $m(L)$ are transferred, using the information given in the rule interface, to edges from $G\backslash m(L)$ to $m(R)$. This defines a rewriting system for higher-order port-graphs. 

Several subgraphs $m(L)$ may exist in $G$ (leading to different rewriting steps); they are computed as solutions of a {\em matching} problem from $L$ to (a subgraph of) $G$. If there is no such injective morphism, we say that $G$ is {\em irreducible} by $L\Ra R$. 

Each {\em rule application} is a rewriting step and a {\em derivation}, or computation, is a sequence of rewriting steps. A port graph on which no rule is applicable is in \emph{normal form}. Rewriting is intrinsically non-deterministic since it may be possible to rewrite several subgraphs of a port graph with different rules or use the same one at different places, possibly obtaining different results.

\section{Automation}
\label{sec:impl}
In this section we give an algorithm to compute the set of all possible rewriting steps from a port-graph $G$, given a set of higher-order rewrite rules $\R$. The extension of the port-graph syntax with higher-order features introduces a potential combinatorial explosion when enumerating all possible rule applications on a given port-graph. This is due to the fact that higher-order variables are matched to  sub-graphs. \todo{example of explosion with disconnected nodes? ex: $m$ higher-ordre nodes in morphism with $n$ nodes} The definition of morphism includes conditions that limit this explosion, especially when dealing with proof port-graphs. 


The algorithm is based on the first-order algorithm implemented in PORGY~\cite{PORGYnotes}, for the original definition of port-graphs. Intuitively, it matches the edges first, identifying the source and target nodes and ports in $G$ and $H$. 
All along the execution of the algorithm, a context is updated, that stores some useful information. The context is seen abstractly as a tuple of:
\begin{itemize}
\item a partial mapping of first-order nodes: $ \pp{image}: V_G \hookrightarrow V_H $
\item a partial mapping of higher-order nodes to sets of nodes: $ \ho{\pp{image}}: \ho V_G \to \Part( V_H \cup \ho V_H ) $
\item a state for each node in $H$ that represents its availability to be matched
\end{itemize}
For instance, adding a node to an image means mapping this node to its image in $\pp{image}$, and putting its state to ``hidden'' so that it cannot be reused in another node's image.

The edge matching is performed using a ``first-order nodes first, constant names first'' priority. This way, the algorithm first provides an image for all the first-order nodes. Some ports in the  images of higher-order nodes are also identified, and used as starting point to match the interface of the corresponding higher-order variables. The algorithm then maps the edges between the other ports of higher-order nodes, and finally maps the  free ports of higher-order nodes
to ports in $H$. In order to enumerate all the solutions, some nodes are added to this image, and to ensure that it defines a proper morphism between $G$ and $H$, check that this image has exactly the same interface as the node.
 

More details  about the two main phases of the algorithm (match the edges and extend the image sets of higher-order nodes) are given in Algorithm~\ref{algo:matching}.

\begin{algorithm}[!h]
  \caption{Matching algorithm.}
  \label{algo:matching}
  \begin{algorithmic}
    \STATE $\bullet$ match the edges between first-order nodes\\
    and update their images accordingly in the context
    \STATE $\bullet$ map all the disconnected first-order nodes
    \STATE $\bullet$ match the edges between first-order nodes and higher-order nodes\\
    setting the first-order nodes' images and adding one node to the higher-order nodes' images in the context
    \STATE $\bullet$ match the edges between higher-order variables\\
    updating the context accordingly
    \STATE $\bullet$ add all the connected nodes to the higher-order images
    \STATE $\bullet$ check the interface of the higher-order images
  \end{algorithmic}
\end{algorithm}

Note that once all edges are matched in Algorithm~\ref{algo:matching}, all first-order nodes, and all higher-order's interface ports are mapped. When every higher--order node has its connected interface mapped, the images of higher-order nodes are extended to greater node sets. The aim is to enumerate all the possible solutions, that is all the tuples of port-graphs of $H$ that constitute valid images for the higher-order nodes in $G$. For this, we find all the possible solutions for the image of the first higher-order node, and for each of these solutions, all the solutions for the second one (that are disjoint with the first ones), and so on\footnote{Even if it seems inefficient, there is no better algorithm in the case where the subject graph has no edges, and all higher-order variables have no interface.}.

For the current definition (where ports of higher-order nodes are variables only), the last check is reduced to count the number of free ports in the sub-graph (free meaning not linked to another port in the same sub-graph). This can be done dynamically, maintaining a variable representing the number of free ports in the image of each higher-order variable. It is easy to extend the interface of higher-order variables to constant and variable node names, and a similar dynamic updating of a list of ports can be performed to achieve the same result then.


 We perform some dynamic checking along the expansion of the higher-order images, to try to prevent the solutions from getting irreversibly wrong (for instance, including nodes that can obviously not be included in higher-order images).


\section{Properties}
\label{sec:prop}

\paragraph{Relating higher-order and first-order port-graphs.}
The higher-order port-graphs defined in this paper constitute a proper extension  of first-order ones. This intuition is reflected by the notation similarities between first- and higher-order, and can be expressed mathematically as follows.

\begin{thm}[Simulation by higher-order variables]
  The solution of the matching problem between two port-graphs $G$ and $H$ is a subset of the solutions of the matching problem of $G'$ by $H$, where $G'$ consists of the graph $G$ where every first-order variable node has been replaced by a higher-order one with the same number of ports. 
More precisely, if for every first-order variable $\fov X_i$, we introduce a higher-order variable $\hov X_i$ with same interface, using the higher-order variable $\hov X_i$ instead of $\fov X_i$ preserves solutions.
\end{thm}

The proof is omitted, but we remark that  the syntax and morphism have been specifically developed with this result in mind.

\paragraph{Specification of proof net and interaction net reductions.}
We briefly present some examples  inspired by~\cite{proof_graphs}, where the original notion of port-graph is used to represent intuitionistic proofs graphically, and to study their normalisation as a rewriting process. In fact, the first-order port-graphs used  in~\cite{proof_graphs} are generalised interaction nets, as indicates the presence of principal ports.

\begin{figure}[h]
  \input{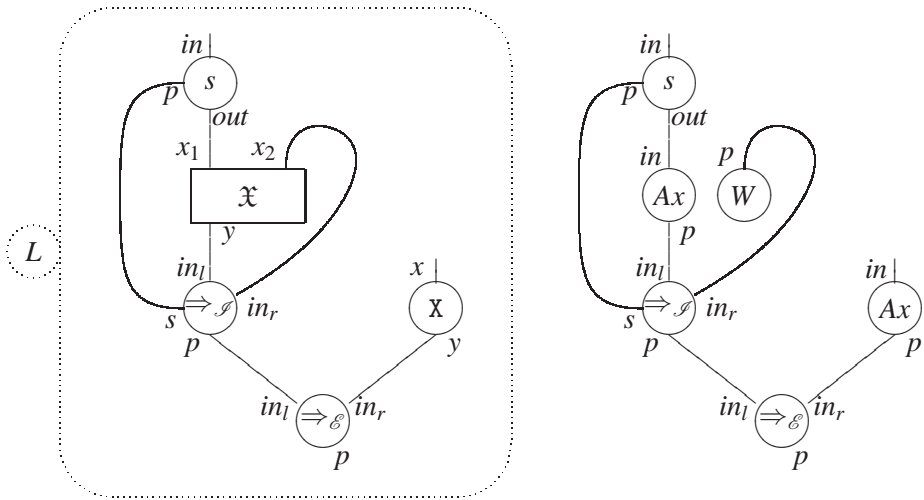}
  \caption{Higher-order pattern and target port-graphs.}
  \label{fig:morphism_ho}
\end{figure}

Figure~\ref{fig:morphism_ho} gives an example of a higher-order pattern $L$, along with a subject graph $G'$. The pattern corresponds to the intuitive formulation of a redex in the cut-elimination procedure (eliminating an introduction of $\Rightarrow$ followed by its elimination). It  is expressed directly with the syntax defined in Section~\ref{sec:hopg}. The higher-order variable $\hov X$ represents a proof. Note that in \cite{proof_graphs}, this single rule was implicitely expanded into a large family of first-order rules to fit the first-order syntax. Although in interaction nets axioms are represented using only edges, here axioms are explicitly represented as nodes. In this way, a higher-order variable with two ports can be mapped to an axiom using the matching algorithm.

\section{Conclusion}
\label{sec:conc}
We have described an extension of the port-graph rewriting notion from \cite{Andrei08,AndreiK08c} with higher-order features, designed to facilitate the modelling of proof normalisation procedures as graph rewriting system.

This extension does not provide more computational power (port graphs are already Turing complete) but if we see port graphs as a specification or modelling tool, the extended language is more expressive in that it allows us more concise, high-level definitions.

Properties of higher-order port-graph rewriting, such as confluence and termination, have not been studied yet.
This will be the subject of future work.


\end{document}